\documentclass{appolb}
\usepackage{epsfig,amssymb,citesort}

\begin{document}
\title{Fragmentation of Partons
\thanks{Presented at XXXIV International Symposium on 
Multiparticle Dynamics\\ July 26-August 1, 2004, Sonoma County, California, USA}
}
\author{Stefan Kretzer
\address{Physics Department and
RIKEN-BNL Research Center,\\ 
Brookhaven National Laboratory, Upton, New York 11973, USA\\
E-mail: Kretzer@bnl.gov}}
\maketitle
\begin{abstract}
The concept of parton fragmentation
in QCD hard scattering phenomenology
as well as 
NLO pQCD analysis of 
fragmentation functions are outlined.
Hadroproduction of pions of a few GeV
$p_\perp$ is discussed through the example of  
recent measurements at $\sqrt{S_{\rm RHIC}}=200$ GeV.
\end{abstract}

\section{Introduction}
Hard pQCD reactions produce highly virtual partons 
that will radiate off their virtualities, thus 
evolving into the non-perturbative states
that contain observable hadrons. A quantitative
understanding of this process, known broadly as 
the fragmentation of partons, considerably
widens the class of reactions that can be handled 
within a perturbative  QCD approach.
Examples are too many to enumerate here 
(as the reader may convince
himself by browsing through the proceedings), 
two recent cases will be discussed below in 
Sections \ref{sec:all} and \ref{sec:rho}.

\begin{figure}[t]
\vspace*{-1.5cm}
\hspace*{1.5cm}
\epsfig{figure=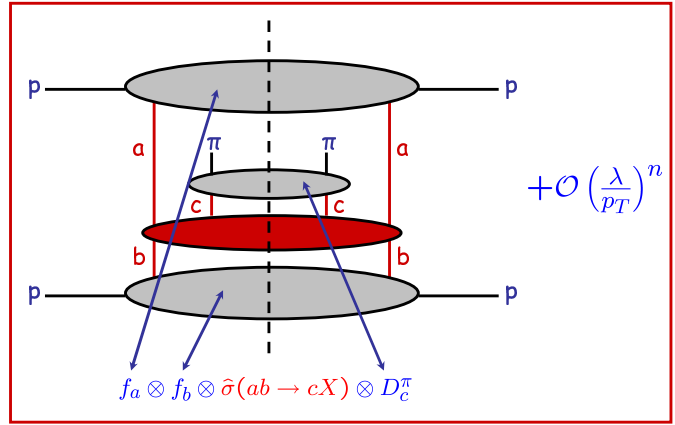,width=0.8\textwidth ,angle=270}
\vspace*{-3.cm}
\caption{Factorization in terms of parton densities 
and fragmentation functions.\label{fig:fac}}
\end{figure}

\begin{figure}[t]
\vspace*{-0.5cm}
\hspace*{-0.5cm}
\epsfig{figure=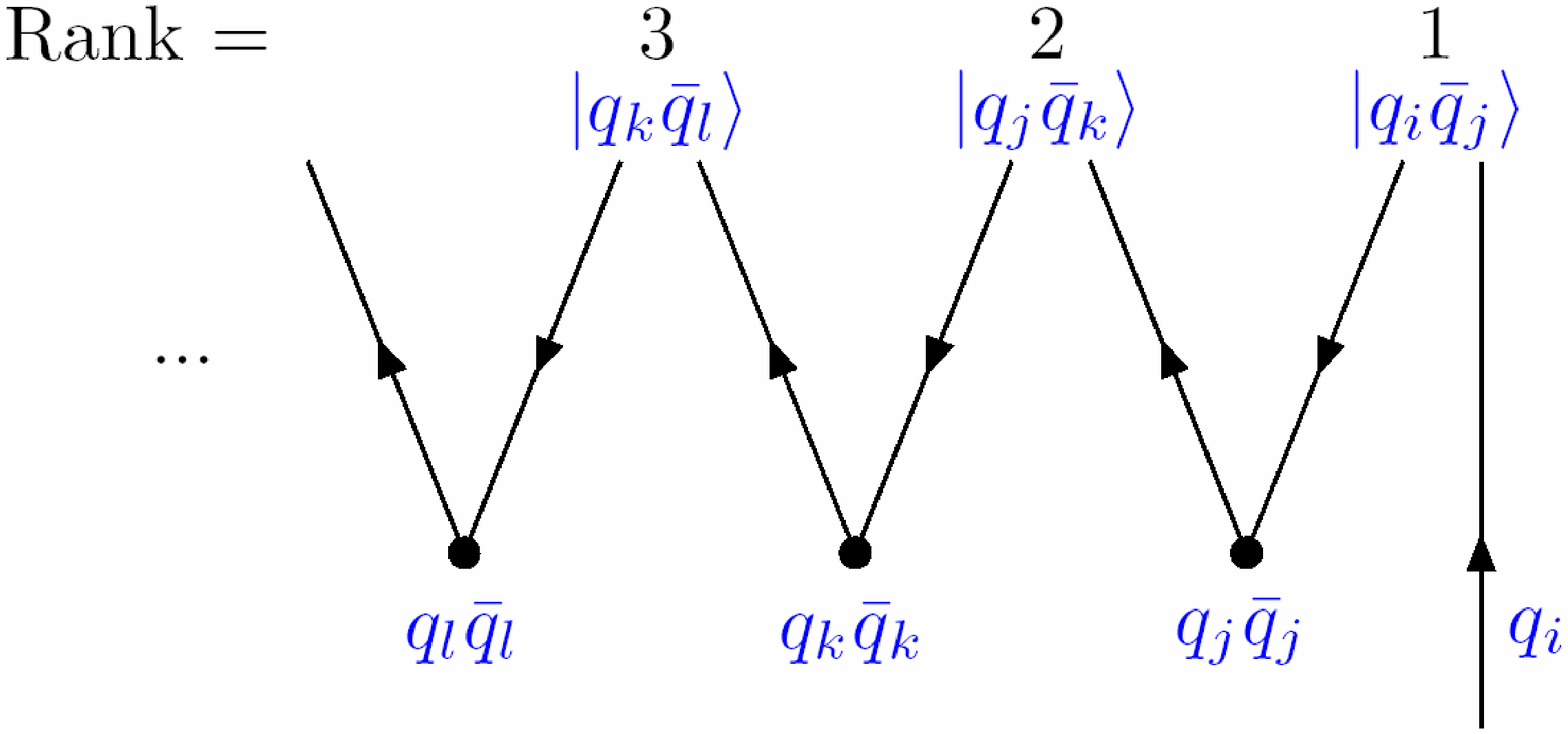,width=0.55\textwidth,angle=270}
\hspace{0.5cm}
\epsfig{figure=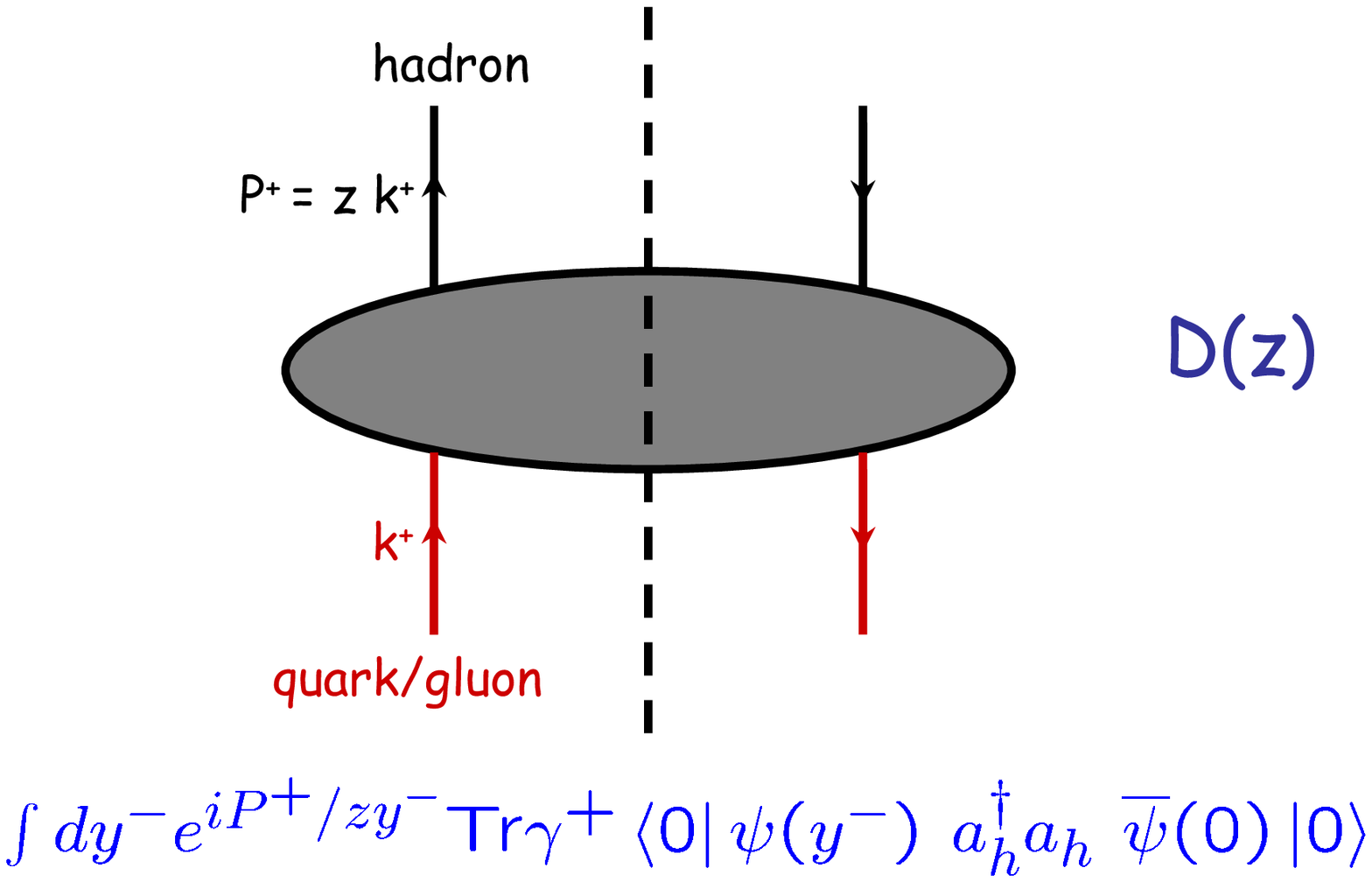,width=0.65\textwidth,angle=270}
\vspace*{-3.cm}
\caption{Field \& Feynman model (left) of cascade fragmentation
and bilocal operator (right) \`{a} la Collins \& Soper.
\label{fig:fiefeycoso}}
\end{figure}

A graph such as in Fig.~\ref{fig:fac} 
serves as a popular illustration
of QCD factorization \cite{fact}.
It also displays the role of fragmentation functions 
$D_c^\pi (z)$ in 
hard scattering phenomenology when there are
detected hadrons in the final state. 
The early Field \& Feynman \cite{fiefey}
cascade model for quark fragmentation into mesons  
is depicted in Fig.~\ref{fig:fiefeycoso} 
side by side with a 
later field-theoretic definition in terms of 
a bilocal operator \cite{cs}. 
At leading twist accuracy, these describe fragmentation
(or decay) functions (now in a narrower sense of 
the word) that turn a single parton into a single-hadron 
inclusive state, with fractional momentum transfer $z$ from
the parton to the hadron. Analogous to parton densities,
renormalization induces evolution of the (transposed) 
DGLAP type:
\begin{equation}\label{eq:dglap}
\frac{\partial D_i}{\partial \ln \mu}
= \sum_{j=q,{\bar q},g} D_j \otimes P_{ij}
\end{equation}

This write-up discusses a few theoretical aspects of
fragmentation functions and fragmentation processes. 
Please confer also K.~Hamacher's contribution to these
proceedings \cite{kh}
for experimental results.

\section{Global Analyses of Fragmentation Functions}
Fragmentation functions are determined in
NLO QCD fits to data. They are statistically dominated by 
$e^+e^-$ annihilations with particularly precise 
$e^+e^- \rightarrow h X$ data from
$Z^0$ decays at LEP and SLD.
Even though the $e^+e^-$ data can be reproduced
very successfully over a wide range of energy,
there are several limitations:
\begin{enumerate}
\item The leading order partonic process  
$e^+e^- \rightarrow q {\bar q}$ produces quarks only;
the subleading terms are too weak to 
constrain gluon fragmentation sufficiently. 
Gluon-quark mixing in the evolution 
Eq.~(\ref{eq:dglap}) is not enough of a constraint, either.
\label{item:gluon}
\item Data are less accurate at large-$z$. This fact translates 
into less accurate functions $\left. D_i(z)\right|_{{\rm large}-z}$.
\label{item:largez} 
\item The individual light flavour channels 
(corresponding to the ``Ranks'' in 
Fig.~\ref{fig:fiefeycoso})
cannot be disentangled
from the flavour inclusive data.
\label{item:flavour}
\end{enumerate}
These items limit, at present, 
the accuracy of the functions $D_i(z)$. 
They require that fragmentation analyses become
truly global the way that analyses of parton densities 
already are, i.e.~including data from different processes 
that constrain different kinematic regions and flavour combinations.

Recent global analyses of fragmentation functions,
determined at Next-to-Leading-Order accuracy in QCD, 
were performed in \cite{ffs}. The individual
fits are conceptually very similar,
such that a comparison between the 
fits gives a good first estimate of the uncertainties in
the functions $D_i(z)$. As explained above,  
the uncertainties of these functions are data driven
and can be large. It can indeed be shown 
\cite{klc} that the electroweak couplings of
up and down type quarks at the $Z^0$ pole suggest that
only flavour-insensitive combinations 
such as the singlet $\sum_q D_{q+{\bar q}}$ are well
determined. Flavour structure can be 
constraint through the
analysis of semi-inclusive 
DIS \cite{klc}. The simple lowest order term
\begin{equation}
d \sigma_{\rm SIDIS} \propto 
\sum_{q} \left[ q(x) D_q(z) 
+{\bar q}(x) D_{\bar q} (z) \right]
\end{equation}
illustrates that the information is local in $z$
(no convolution at LO) and that
the by now very good knowledge of the
parton distributions $q(x)$ provides reliable flavour weights.

Having discussed item \ref{item:flavour},
we are still  left with items \ref{item:gluon}, \ref{item:largez}
from our shopping list above. They lead over to the analysis
of hadroproduction reactions. 

\section{Fragmentation in Hadroproduction Reactions}

\subsection{Partonometry at $\sqrt{S_{\rm RHIC}}=200$ GeV}
RHIC ($pp$ mode) operates at $\sqrt{S_{\rm RHIC}}=200$ GeV, 
between lower energies 
(such as ISR or Fermilab), where the pQCD calculations are
known to be problematic \cite{bs}, and higher CMS energies
(up to $\sqrt{S_{\rm Tevatron}}$), where
pQCD seems to work \cite{ffs}.
It was quite remarkable to see, how well pion production
data in RHIC $pp$ collisions \cite{ernst}
agree with the NLO QCD calculation 
\cite{JSSV}
at central and
forward rapidity and down to $p_\perp$ values as low as
$p_\perp \gtrsim 1$ GeV. While it remains a challenge to understand
the $\sqrt{S}$ dependence
of hadroproduction spectra in terms of power corrections 
\cite{ht}, 
the good agreement of the pion spectrum 
with NLO QCD at  $\sqrt{S_{\rm RHIC}}$ and the good precision of
the RHIC data are a goldmine for fragmentation analyses.

\begin{figure}[t]
\vspace*{-2.5cm}
\hspace*{-1.5cm}
\epsfig{figure=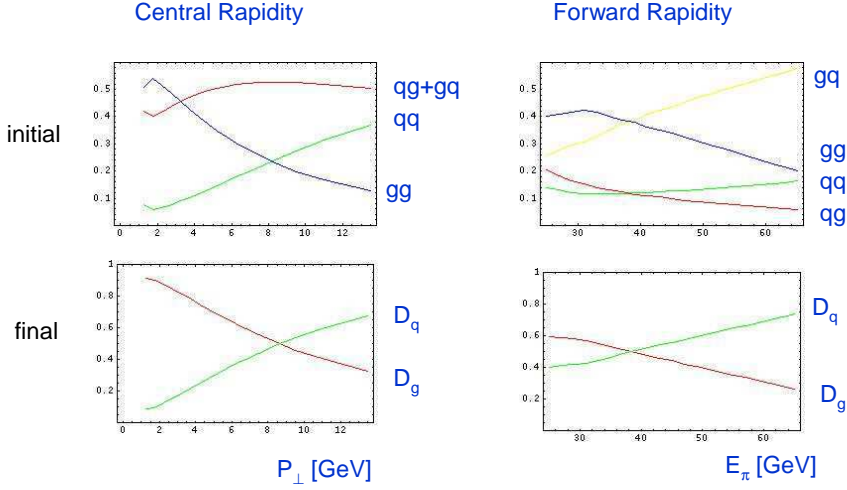,width=1.1\textwidth,angle=270}
\vspace*{-5.cm}
\caption{Partonic decomposition of the initial and final state
of $pp$ collisions at 200 GeV; central and forward rapidity.
\label{fig:fracs}}
\end{figure}

\begin{figure}[t]
\vspace*{-4.5cm}
\hspace*{-2.5cm}
\epsfig{figure=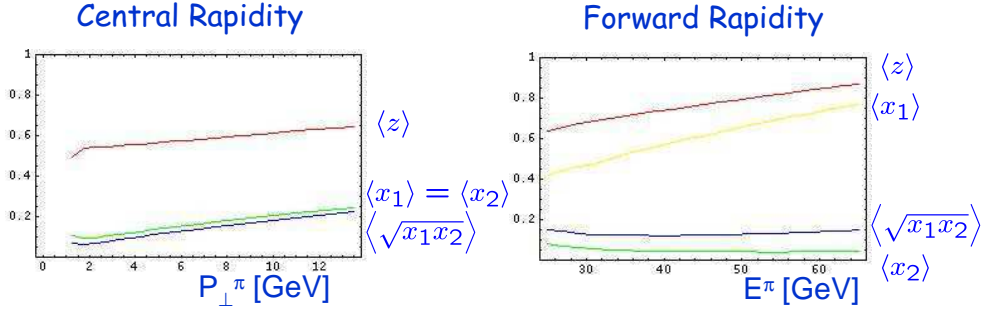,width=1.4\textwidth,angle=270}
\vspace*{-8.5cm}
\caption{Scaling variables, averaged over the parton 
initial and final state
of $pp$ collisions at 200 GeV; central and forward rapidity.
\label{fig:avxz}}
\end{figure}

Fig.~\ref{fig:fracs} decomposes
RHIC collisions into 2-parton initial and 
1-parton-inclusive final states. For Fig.~\ref{fig:avxz}
the corresponding scaling variables have been evaluated 
at their average values, manifesting the symmetric
and asymmetric kinematics of central and forward
rapidities, respectively.
The parton densities are probed in regions
of $x_{1,2}$
where they are well known already 
whereas new constraints are imposed on the
large-$z$ fragmentation functions, and in particular on the
gluon function. While the constraint on the gluon 
fragmentation function is nonlocal (convoluted) in
hadroproduction reactions, a QCD analysis of this reaction is more
straightforward than an analysis of anti-tagged
gluon jets in $e^+e^-\rightarrow b {\bar b} g$ 
configurations \cite{kh}. The latter
are nontrivial due to the fact that jets and fragmentation functions 
are not the same in QCD.

I will next leave the fragmentation functions
and move on to discussing theoretical aspects of two recent 
pion production measurements at RHIC.

\subsection{The double-spin asymmetry $A_{LL}^{\pi}$}
\label{sec:all}
The gluon helicity contribution 
$\Delta g$
to the proton spin is
probed in collisions of longitudinally polarized protons 
under the condition that the dominant dynamics are 
perturbative and of leading twist origin. 
These conditions may apply to the measurement
of beam helicity asymmetries 
\begin{equation}\label{eq:asydef}
A_{\rm{LL}}=\frac{d\Delta \sigma}{d\sigma}=
\frac{d\sigma^{++} - d\sigma^{+-}}{d\sigma^{++} + d\sigma^{+-}}
\end{equation}
for high transverse momentum reactions 
where $d \sigma \equiv 
\left. d \sigma / dp_\perp\right|_{{\rm large}-p_\perp}$. 
For large $p_\perp$ pion production,
a variant of Fig.~\ref{fig:fac} with polarized beams
illustrates that the (glue-glue contribution to the)
reaction
can schematically be written as 
\begin{equation}
A_{\rm{LL}} \propto 
\left( \Delta g_a \otimes \Delta g_b \right)
\otimes \Delta {\hat \sigma}^{ab\rightarrow cX}
\otimes D_c^{\pi}\ .
\label{eq:DDsD}
\end{equation}
The Mellin transform
\begin{equation}
\Delta g_a (N) \Delta g_b(N) {\hat \sigma}^{ab\rightarrow cX}(N) 
D_c^{\pi} (N)
\end{equation}
[with $\Delta g_a(N) \equiv \int_0^1 x^{N-1} g_a(x)$ 
etc.] turns the nonlocal (in parton momentum, 
i.e.~$x_a\neq x_b$) 
convolutions
\begin{equation}\label{eq:conv}
\Delta g_a \otimes g_a \dots
= \int_{x_T^2}^1 dx_a  
\int_{x_T^2/x_a}^1 dx_b\ \Delta g_a(x_a) \Delta g_b(x_b) \ \dots
\end{equation}
into products that are local in the moments $N$. 
For small to medium $p_\perp \lesssim 4$ GeV one can then 
show \cite{prl} by variation
$
{\delta A_{\rm{LL}}}/{\delta \Delta g} =0
$
that $A_{LL}^{\pi}$ has a minimum and 
is basically positive definite
at leading twist, i.e.
\begin{equation}
\left. A_{LL}^{\pi}\right|_{p_\perp \lesssim 4\ {\rm GeV}}
> {\cal{O}} \left( -10^{-3} \right)
\end{equation}
is a necessary condition for leading twist dynamics 
to be dominant in the
production of 
small to medium $p_\perp$ pions. 
To achieve a {\it slightly} negative $A_{LL}^{\pi}$ of
order ${\cal{O}} \left( -10^{-3} \right)$ the Mellin
inversion of $\Delta g_{a,b}(N)$ back to  $\Delta g_{a,b}(x_{a,b})$
has to generate a crossing of   $\Delta g(x)$ through zero and
explore the fact that the convolution
in Eq.~(\ref{eq:conv}) can then pick $x_a\neq x_b$
such that  $\Delta g_{a}(x_{a})\Delta g_{b}(x_{b}) <0$.
For any foreseeable experimental accuracy
 ${\cal{O}} \left( -10^{-3} \right)\simeq 0$ which is why
$A_{LL}^\pi$ is ``basically'' positive definite.
At present, data \cite{phenix} 
are statistically not conclusive  
and the positivity of $A_{LL}^{\pi}$ remains yet to 
be confirmed, or disproved.

\subsection{Production of $\rho (\rightarrow 2 \pi)$ mesons}
\label{sec:rho}
Recent data \cite{star} indicate that the invariant mass spectrum
of pion pairs in $pp$ collisions peaks 
at a value around, but significantly below,
the $\rho$ mass as measured in $e^+e^-$ exclusive production. 
Now, resonance mass shifts would seem an unlikely manifestation
of a perturbative effect if it were not for the fact
that the effect extends to 
large values of transverse momentum $p_\perp \sim 3$ GeV where
the $\rho$ can be expected to be a parton fragment. The strong
decay $\rho \rightarrow 2 \pi$ then contributes a p-wave resonance
to the $D_{i}^{2 \pi}$ fragmentation function
\cite{br}.\footnote{For
the semi-quantitative
argument we are making here, we will ignore
the mixing of $m$-hadron fragmentation with $n$-hadron 
($n<m$) fragmentation, as well as the non-resonant $s$ wave channel.}

In Fig.~\ref{fig:fac},
the fragmenting parton momentum $k$ 
runs over the 
following virtualities and transverse momenta:
\begin{equation}
\int_{M^2/z}^{\hat{s}} 
d k^2\ \int_0^{(1-z)/z(k^2-M^2/z)} d k_{\perp}^2\ \ \ .
\label{eq:kps}
\end{equation}
Strictly, Eq.~(\ref{eq:kps}) differs from the
operator product depicted 
in Fig.~\ref{fig:fiefeycoso} which is defined
to be UV-divergent [after integration over 
$\int d^4 k \delta (k^+ - P^+/z)$].
The argument I am about to make, though, 
is not related to  
the subtle \cite{col03} problems that arise from cutting off 
momentum integrals. 

The partonic center of mass energy ${\hat s}$ can, for an
order of magnitude estimate, be inferred from Fig.~\ref{fig:avxz}
through $\left< {\hat s}\right>=\left< x_1 x_2\right> S_{\rm RHIC}$.
Under the assumption that the partonic spectrum behaves like
$k^{-2}$, a Breit-Wigner
peak in the $2 \pi$ mass spectrum -- positioned at
the $\rho$ mass for $p_\perp \rightarrow \infty $ --
is then shifted to a lower value by a few 10 MeV around
a finite $p_\perp=3\ {\rm GeV}$. 
This suggests a partonic effect that is dual to the pion phase
space distortion at low $p_\perp \rightarrow 0$.
A more detailed evaluation will be given in \cite{bdkr}.

\section{Conclusions}
Parton fragmentation processes explore conceptually and
phenomenologically rich aspects of
hadronization, as I hope to 
have demonstrated through the examples of a positivity bound on the 
spin asymmetry $A_{LL}^\pi$
and a perturbative ansatz to explain resonance mass
shifts at high $p_\perp$.
I have also outlined the next steps in
global analysis of fragmentation functions; a corresponding
update will be 
made available very soon.

\section{Acknowledgments}
Many thanks to M.~Grosse-Perdekamp for the invitation and
to all organizers of ISMD04, in particular 
B.~Gary, for financial support and a superbly organized workshop.
This write-up benefited from
discussions and correspondences with A.~Bacchetta,
K.~Barish, S.~Brodsky, G.~Bunce, P.~Fachini, 
K.~Hamacher, E.~Laenen, A.~Mueller,
M.~Radici, E.~Sichtermann, R.~Seto, and W.~Vogelsang.
Work supported by  RIKEN-BNL and the U.S.~Department 
of Energy (contract DE-AC02-98CH10886).

\end{document}